\documentclass[twocolumn,showpacs]{revtex4}
\usepackage{graphicx}% Include figure files
\usepackage{bm}% bold math
\usepackage{amsmath}

\begin{document}
%\draft
%%%%%%%%%

%%%%%%%%%%%%%%%%%%
%%% Title Page %%%
%%%%%%%%%%%%%%%%%%
%%% Title & Authors %%%
\title{Polarization phenomena in hyperon-nucleon scattering}
\author{S. Ishikawa} \email[E-mail:]{ishikawa@i.hosei.ac.jp}
\author{M. Tanifuji}
\affiliation{Department of Physics, Science Research Center, Hosei University, 
2-17-1 Fujimi, Chiyoda, Tokyo 102-8160, Japan}
\author{Y. Iseri}
\affiliation{Department of Physics, Chiba-Keizai College, 
4-3-30 Todoroki-cho, Inage, Chiba 263-0021, Japan}
\author{Y. Yamamoto}
\affiliation{Physics Section, Tsuru University, Yamanashi 402-8555, Japan}

\date{\today}

\begin{abstract}
We investigate polarization observables in hyperon-nucleon scattering by decomposing scattering amplitudes into spin-space tensors, where each component describes scattering by corresponding spin-dependent interactions, so that contributions of the interactions in the observables are individually identified.
In this way, for elastic scattering we find some linear combinations of the observables sensitive to particular spin-dependent interactions such as  symmetric spin-orbit (LS) interactions and antisymmetric LS ones. 
These will be useful to criticize theoretical predictions of the interactions when the relevant observables are measured.  
We treat vector analyzing powers, depolarizations, and coefficients of polarization transfers and spin correlations, a part of which is numerically examined in $\Sigma^{+} p$ scattering as an example. 
Total cross sections are studied for polarized beams and targets as well as for unpolarized ones to investigate spin dependence of imaginary parts of forward scattering amplitudes. 
\end{abstract}

% 21.45.+v Few-body systems
% 21.30.-x Nuclear forces (see also 13.75.C 
%	  Nucleon-nucleon interactions)
% 21.80.+a Hypernuclei
% 24.70.+s Polarization phenomena in reactions
%
% 25.80.-e Meson- and hyperon-induced reactions  
% 25.80.Pw Hyperon-induced reactions  

% 13.75.Ev Hyperon-nucleon interactions  
%

\maketitle

%%%%%%%%%%%%%%%%%%%%%%
\section{Introduction}
\label{Sec:I}

Interactions between hyperons and nucleons are fundamental subjects in studies of nuclear structures and reactions that contain hyperons.  
So far, a number of theoretical models for hyperon-nucleon ($YN$) interactions have been developed based on boson-exchange models \cite{Ma89,Ri99,Ho89} or quark-cluster ones \cite{Fu96,Ta00}. 
On the other hand, experimental studies of the interactions through YN scattering have scarcely been performed, which leaves many ambiguities particularly on their spin dependence. 

As is well known, polarization phenomena are a substantial tool in studies of spin-dependent interactions. 
In a phase-shift analysis of $YN$ scattering \cite{Na02}, which is an information source of the spin dependent interactions, polarization observables are shown to be indispensable to avoid ambiguities. 

Recently, asymmetries of scattered hyperons have been measured for elastic scattering of polarized $\Sigma^{+}$ and $\Lambda$ on protons \cite{Ka02}, which provide an experimental evidence on characteristic difference of the spin dependence between ${\Sigma}^{+} p$ and $\Lambda p$ interactions, though of qualitative nature at present. 
Considering that such kinds of experimental research will be developed more in the future, we will theoretically investigate polarization observables in $YN$ scattering in relation to the spin dependence of $YN$ interactions. 
In general, contributions from different kinds of spin-dependent interactions are mixed up in scattering observables.
However, we will predict some linear combinations of the observables to exhibit effects of particular spin-dependent interactions. 
Analyses of such combinations will thereby be useful to clarify characteristics of the interactions and provide clear-cut criticisms on the spin dependence of the model interactions when the relevant observables are measured.  

In order to relate the polarization observables to the spin-dependent interactions, we will decompose the scattering amplitudes into scalar, vector, etc., in the spin space,  each of which describes scattering by central interactions, by spin-vector interactions like LS ones, etc., respectively.
When the observables are described in terms of such amplitudes, one will be able to identify the contributions of particular spin-dependent interaction in the observables accordingly.  
Similar decompositions of the scattering amplitudes have been applied to analyses of nucleon-deuteron scattering \cite{Is01,Is02,Is03}, which have provided deeper understanding for effects of spin-dependent interactions in the scattering observables and have succeeded in clarifying the scalar, vector, and tensor characters of three-nucleon forces. 
Such success encourages us to extend the method to the $YN$ scattering.  

In the present paper, we will consider a general case of scattering between two spin-1/2 particles, since the spins of nucleons and hyperons, $\Lambda$, $\Sigma$, etc., are 1/2.  
In Sec.\ \ref{Sec:II}, the decomposition of the scattering amplitude into the spin-space tensors is given in a model independent way. 
Each component of the decomposed amplitude is related to conventional amplitudes by giving explicit forms for the tensors. 
In this way, elastic scattering is investigated in detail.
As will be shown later, the present amplitude includes a vector component effective for mixing of  total intrinsic spins, which is lacked in the amplitude of nucleon-nucleon scattering.  
Thus the polarization observables in the $YN$ scattering are composed of the constituents in a way different from those in the nucleon-nucleon scattering \cite{Ke59}.
Using the amplitudes given in Sec.\ \ref{Sec:II}, we investigate typical polarization observables for elastic scattering in Sec.\ \ref{Sec:III}, where the analyzing powers and second order polarization observables such as  depolarizations are treated.  
Further, it is shown that total cross sections for the polarized beam and target as well as for the unpolarized ones exhibit contributions of particular spin-dependent interactions to the scattering amplitudes, when linear combinations are considered. 
In Sec. \ref{Sec:IV}, a part of theoretical predictions is numerically examined as an example for the $\Sigma^{+} p$ scattering by using the Nijmegen interactions \cite{Ri99}, where the calculated quantities are compared between different versions of the interactions. 
Summary will be given in Sec.\ \ref{Sec:V}.
%Explicit forms for the relation between the scattering amplitudes and the decomposed amplitudes are presented in Appendix \ref{Sec:AppA}, 
%some details on the time reversal theorem are given in Appendix \ref{Sec:AppB}, 
%and general expressions of the second order polarization observables are summarized in Appendix \ref{Sec:AppC}.

%%%%%%%%%%%%%%%%%%%%%%%%%%%%%%%%%%%%%%%%%%%%%%%%%%%%%%%%%%%%%%%%%
\section{Scattering amplitudes for two spin-1/2 particles}
\label{Sec:II}

%%%%%%%%%%%%%%%%%%%%%%%%%%%%%%%
\subsection{Spin Tensor Analysis of Scattering Amplitudes}

Let us consider the T-matrix $\bm{M}$ for scattering of two spin-1/2 particles, $a+b \to c+d$, characterized by the isospin and the strangeness, where the parity is conserved. 
The matrix element of $\bm{M}$ gives the scattering amplitude as usual.
To decompose the amplitude according to the tensorial property in the spin space, we will expand $\bm{M}$ by spin-space tensors of the rank $K$ and $z$ component $\kappa$, $\bm{S}^{(K)}_{\kappa}$, 
\begin{equation}
  \bm{M}
  =\sum_{K \kappa} (-)^{\kappa} \bm{S}^{(K)}_{-\kappa}
     \bm{R}^{(K)}_{\kappa},
\label{eq:MSK}
\end{equation}
where $\bm{R}^{(K)}_{\kappa}$ is a coordinate-space tensor associated with $\bm{S}^{(K)}_{-\kappa}$. 
Then the matrix element of $\bm{M}$ designated by the $z$ components of spins of the related particles, $\nu_a$  etc., and the relative momenta between the particles in the initial and final states, $\bm{k}_{\textrm{i}}$ and $\bm{k}_{\textrm{f}}$, is given by
\begin{eqnarray}
&& \langle \nu_c \nu_d; \bm{k}_{\textrm{f}} | \bm{M} | \nu_a \nu_b; \bm{k}_{\textrm{i}} \rangle
=  \sum_{s_{\textrm{i}} \nu_{\textrm{i}} s_{\textrm{f}} \nu_{\textrm{f}}} (\frac12 \frac12 \nu_a \nu_b|s_{\textrm{i}} \nu_{\textrm{i}})
\nonumber\\
&& \times  (\frac12 \frac12 \nu_c \nu_d|s_{\textrm{f}} \nu_{\textrm{f}}) 
      \sum_{K\kappa} (-)^{s_{\textrm{f}}-\nu_{\textrm{f}}} (s_{\textrm{i}} s_{\textrm{f}} \nu_{\textrm{i}} -\nu_{\textrm{f}} |K\kappa)
\nonumber\\
&& \times
  M^{(K)}_{\kappa}(s_{\textrm{i}} s_{\textrm{f}};\bm{k}_{\textrm{i}} \bm{k}_{\textrm{f}}),
\label{eq:Melemnts}
\end{eqnarray}
where the geometrical part of the matrix element of $\bm{S}_{\kappa}^{(K)}$ is described by the Clebsch-Gordan coefficient due to the Wigner-Eckart theorem, and the physical part is included in the last factor $M^{(K)}_{\kappa}(s_{\textrm{i}} s_{\textrm{f}}; \bm{k}_{\textrm{i}} \bm{k}_{\textrm{f}})$, which is an amplitude of rank $K$ and is given by
\begin{eqnarray}
M^{(K)}_{\kappa}(s_{\textrm{i}} s_{\textrm{f}}; \bm{k}_{\textrm{i}} \bm{k}_{\textrm{f}}) &=& 
  \frac{(-)^{s_{\textrm{i}}-s_{\textrm{f}}}}{\sqrt{2K+1}} (s_{\textrm{f}}||\bm{S}^{(K)}||s_{\textrm{i}})
\nonumber\\
&& \times \langle\bm{k}_{\textrm{f}}|\bm{R}^{(K)}_{\kappa}|{\bm{k}}_{\textrm{i}} \rangle.
\label{eq:M_Kkappa}
\end{eqnarray}

In the choice of the Madison Convention for the reference axes, $\hat{\bm{z}}||\bm{k}_{\textrm{i}}$ and $\hat{\bm{y}}||\bm{k}_{\textrm{i}} \times \bm{k}_{\textrm{f}}$, $\bm{R}_{-\kappa}^{(K)}$ is related to $\bm{R}_{\kappa}^{(K)}$ due to the parity conservation \cite{Ma70} as 
\begin{equation}
\bm{R}_{-\kappa}^{(K)} = (-)^{K-\kappa} \bm{R}_{\kappa}^{(K)}, 
\label{eq:RK}
\end{equation}
which leads to
\begin{equation}
M_{-\kappa}^{(K)} (s_{\textrm{i}} s_{\textrm{f}};\bm{k}_{\textrm{i}}  \bm{k}_{\textrm{f}}) 
  = (-)^{K-\kappa}M_{\kappa}^{(K)}(s_{\textrm{i}}s_{\textrm{f}}; \bm{k}_{\textrm{i}}  \bm{k}_{\textrm{f}}).
\end{equation}
Thus the scattering amplitude consists of the following non-vanishing independent amplitudes classified by the rank of the spin-space tensor: 
the scalar amplitudes $U_j$ $(j=0,1)$
\begin{equation}
U_j= M^{(0)}_{0}(j j; \bm{k}_{\textrm{i}} \bm{k}_{\textrm{f}}),
\label{eq:Uj}
\end{equation}
the vector ones $S_j$ $(j=1, 2, 3)$
\begin{subequations}
\label{eq:Sj}
\begin{eqnarray}
S_1&=& M^{(1)}_{1}(0 1; \bm{k}_{\textrm{i}} \bm{k}_{\textrm{f}}),
\\
S_2&=& M^{(1)}_{1}(1 0; \bm{k}_{\textrm{i}} \bm{k}_{\textrm{f}}),
\\
S_3&=& M^{(1)}_{1}(1 1; \bm{k}_{\textrm{i}} \bm{k}_{\textrm{f}}),
\end{eqnarray}
\end{subequations}
and the tensor ones $T_j$ $(j=1, 2, 3)$
\begin{equation}
T_j = M^{(2)}_{j-1}(1 1; \bm{k}_{\textrm{i}} \bm{k}_{\textrm{f}}).
\label{eq:Tj}
\end{equation}
These amplitudes describe the scattering by interactions with the corresponding tensor property, where contributions of higher orders of the interactions are included under the restriction due to the tensorial property. 
For example, the scalar amplitude $U_j$'s include the higher order contributions  as long as they form scalars in the spin space.

The present scattering amplitude is equivalent to the Wolfenstein amplitude in Ref.\ \cite{Ho89} in the sense that both are composed of two scalar components, three vector ones, and three tensor ones. 
Also, such decomposition of the scattering amplitude into spin-space tensor components is based on the theoretical development in Ref.\ \cite{Ta68} and is similar to that in Ref.\ \cite{Cr02} for nucleon-nucleon inelastic scattering. 
In practical cases, $U_j$, $S_j$, and $T_j$ are calculated from $\langle \nu_c \nu_d; \bm{k}_{\textrm{f}} | \bm{M} | \nu_a \nu_b; \bm{k}_{\textrm{i}} \rangle$, which will be obtained in conventional ways.
More details are given in Appendix \ref{Sec:AppA}.

In the elastic scattering, time-reversed states are equivalent to the original ones. 
Then applying the time-reversal theorem \cite{Go64} to the matrix element of $\bm{M}$, we get
\begin{eqnarray}
&&\langle \nu_c \nu_d ; \bm{k}_{\textrm{f}} | \bm{M} | \nu_a \nu_b ; \bm{k}_{\textrm{i}} \rangle = (-)^{\nu_c+\nu_d-\nu_a-\nu_b}
\nonumber\\
 && \times
  \langle -\nu_a -\nu_b; -\bm{k}_{\textrm{i}} | \bm{M} | -\nu_c -\nu_d ;-\bm{k}_{\textrm{f}} \rangle,
\end{eqnarray}
which leads to
\begin{eqnarray}
& &M^{(K)}_{\kappa}(s_{\textrm{i}} s_{\textrm{f}}; \bm{k}_{\textrm{i}} \bm{k}_{\textrm{f}}) 
\nonumber\\
&=& (-)^{s_{\textrm{i}}+s_{\textrm{f}}-K} 
   M^{(K)}_\kappa(s_{\textrm{f}} s_{\textrm{i}};  -\bm{k}_{\textrm{f}} -\bm{k}_{\textrm{i}}).
\end{eqnarray}
This gives the following relations for the vector and tensor amplitudes, the derivation of which is given in Appendix \ref{Sec:AppB}, 
\begin{equation}
S_2=-S_1 
\label{eq:S_TimeRev}
\end{equation}
and 
\begin{equation}
\frac12 \sin \theta \left(\sqrt{\frac32} T_1-T_3 \right) =
   -\cos\theta T_2.
\label{eq:T_TimeRev}
\end{equation}

Thus the independent amplitudes for the elastic scattering are two scalar ones, two vector ones, and two tensor ones. 
Since the composition of independent amplitudes depends on the property of the scattering, we will specify the scattering to the elastic one for further developments. 
%For example in $\bar{p} p \to \bar{\Lambda} \Lambda$ reactions \cite{Ke00}, the number of the independent amplitudes is also six but they are composed of two scalar amplitudes, one vector one and three tensor ones due to the pair annihilation and creation mechanisms. 

%%%%%%%%%%%%%%%%%%%%%%%%%%%%%%%%%%%%
\subsection{Conventional Representation for Elastic Scattering}

For the elastic scattering, $a + b \to a + b$, the T-matrix will be represented in terms of spin-independent, spin-spin, symmetric LS (SLS), antisymmetric LS (ALS), and tensor components as
\begin{eqnarray}
\bm{M} &=& V_{\textrm{c}} +V_\sigma \left( \bm{s}_a \cdot \bm{s}_b \right) 
\nonumber\\
&&  + V_{\textrm{SLS}} \left(\bm{s}_a + \bm{s}_b \right) \cdot \bm{L}  
  + V_{\textrm{ALS}} \left(\bm{s}_a - \bm{s}_b \right) \cdot \bm{L} 
\nonumber\\
&&    + V_{\textrm{T}} \left([ \bm{s}_a \otimes \bm{s}_b]^{(2)} \cdot \bm{Y}_{2}(\hat{\bm{r}}) \right),
\label{eq:M_conv}
\end{eqnarray}
where $\bm{L}$ is the $a$-$b$ relative orbital angular momentum, $\bm{r}$ is the $a$-$b$ relative coordinate, and 
$V^{,}$s are form-factor functions, which include the higher order effects, for  the spin-independent central interaction ($V_{\textrm{c}}$), the spin-spin interaction ($V_\sigma$), the SLS interaction ($V_{\textrm{SLS}}$), 
 the ALS interaction ($V_{\textrm{ALS}}$), and the tensor interaction ($V_{\textrm{T}}$). 
Here, exchange effects due to strangeness transfers between the particles are included in $V_{\textrm{ALS}}$. 
For the nucleon-nucleon scattering, the term of $V_{\textrm{ALS}}$ is eliminated because of the equivalence of $a$ and $b$.

We will connect the amplitudes of Eqs.\ (\ref{eq:Uj})-(\ref{eq:Tj}) to those in Eq.\ (\ref{eq:M_conv}) by specifying $\bm{S}^{(K)}$ and $\bm{R}^{(K)}_{\kappa}$ in Eq.\ (\ref{eq:M_Kkappa}) as $\bm{1}$ and $V_{\textrm{c}}$, $(\bm{s}_a \cdot \bm{s}_b)$ and $V_\sigma$, $\bm{s}_{a} \pm \bm{s}_{b}$ and $V_{\textrm{SLS}} L_{\kappa=1}$ ($V_{\textrm{ALS}} L_{\kappa=1}$), where the terms of $\kappa=1$ are effective due to Eq.\ (\ref{eq:Sj}), etc.
For this purpose, we define new scalar amplitudes, 
\begin{subequations}
\label{eq:UaUb}
\begin{eqnarray}
U_\alpha &\equiv& \langle \bm{k}_{\textrm{f}} |V_{\textrm{c}}| \bm{k}_{\textrm{i}} \rangle, 
\\
U_\beta &\equiv& \langle \bm{k}_{\textrm{f}} |V_\sigma| \bm{k}_{\textrm{i}}\rangle,
\end{eqnarray}
\end{subequations}
and new vector ones, 
\begin{subequations}
\label{eq:SaSb}
\begin{eqnarray}
S_{\alpha} &\equiv& \langle \bm{k}_{\textrm{f}} |V_{\textrm{ALS}} L_{1}| \bm{k}_{\textrm{i}} \rangle,
\\
S_{\beta} &\equiv& \langle \bm{k}_{\textrm{f}} |V_{\textrm{SLS}} L_{1}| \bm{k}_{\textrm{i}}\rangle,
\end{eqnarray}
\end{subequations}
and obtain
\begin{subequations}
\begin{eqnarray}
U_0 &=& U_{\alpha}-\frac34 U_{\beta}, 
\\
U_1 &=& \sqrt3 (U_{\alpha} +\frac14  U_{\beta}),
\end{eqnarray}
\end{subequations}
and 
\begin{subequations}
\begin{eqnarray}
S_1 &=& -S_2 = -S_{\alpha}, 
\\
S_3 &=& {\sqrt2} S_{\beta}.
\label{eq:S3Sab}
\end{eqnarray}
\end{subequations}
Then $S_1(=-S_2)$ describes the scattering by the ALS interaction and $S_3$ the one by the SLS interaction. 
The former interaction couples the states of the total intrinsic spins 0 and 1, while the latter interaction does not. 

The tensor amplitudes $T_j$ $(j=1, 2, 3)$ are calculated as
\begin{equation}
T_j = \frac12 \langle \bm{k}_{\textrm{f}} |V_{\textrm{T}} Y_{2, j-1}|\bm{k}_{\textrm{i}}\rangle,
\label{eq28}
\end{equation}
where one of $T_j$ is not independent due to the time reversal theorem, Eq.\ (\ref{eq:T_TimeRev}). 
For later convenience, we will choose independent amplitudes $T_{\alpha}$ and $T_{\beta}$ as
\begin{subequations}
\label{eq:TaTb}
\begin{eqnarray}
T_{\alpha}&=&\frac1{\sqrt6} T_1 +T_3, 
\\
T_{\beta} &=& \frac1{\sqrt6} T_1 -T_3,
\end{eqnarray}
\end{subequations}
which give
\begin{equation}
T_2=-\tan\theta (\frac12T_{\alpha} +T_{\beta}).
\label{eq100Y}
\end{equation}

%%%%%%%%%%%%%%%%%%%%%%%%%%%%%%%%%%%%%%%%%%%%%%%%%%%%%%%%%
\section{Polarization observables}  %in elastic scattering}
\label{Sec:III}
%%%%%%%%%%%%%%%%%%%%%%%%%%%%%%%%%%%%%%%%%%%%%%%%%%%%%%%%%%%

In this section, we will calculate analyzing powers, depolarizations, polarization transfer coefficients, and spin correlation coefficients for the elastic scattering, $a + b \to a+ b$, and total cross sections using the scattering amplitudes derived in the preceding section and show their linear combinations sensitive to individuals of the scalar, vector, and tensor interactions.

%%%%%%%%%%%%%%%%%%%%%%%%%%%%%%%%%%%%
\subsection{Vector analyzing powers}

The vector analyzing powers $A_y(a)$ for the polarized beam $a$ and $A_y(b)$ for the polarized target $b$, which are equivalent to respective cross-section asymmetries, are defined as
\begin{subequations}
\begin{eqnarray}
A_y(a) &=& \frac1{N_R} \textrm{Tr} \left(\bm{M} \sigma_y(a) \bm{M}^\dag\right), 
\\
A_y(b) &=& \frac1{N_R} \textrm{Tr} \left(\bm{M} \sigma_y(b) \bm{M}^\dag\right),
\end{eqnarray}
\end{subequations}
where $N_R$ is given by
\begin{eqnarray}
N_R &\equiv& \textrm{Tr} \left(\bm{M} \bm{M}^\dag \right)
\nonumber\\
 &=&|U_0|^2+|U_1|^2 + 2 \left(|S_1|^2 +|S_2|^2 +|S_3|^2 \right)
\nonumber\\
 & & + |T_1|^2 + 2 \left( |T_2|^2 +|T_3|^2\right)
\label{eq:NR_1}
\end{eqnarray}
and is related to differential cross sections $d\sigma/d\cos\theta$ as
\begin{equation}
\frac{d\sigma}{d\cos\theta} = \frac{2\pi k_{\textrm{f}}}{4k_{\textrm{i}}} N_R.
\label{eq:Difcs}
\end{equation}

Using the amplitudes in Eqs.\ (\ref{eq:Uj})-(\ref{eq:Tj}), we get
\begin{subequations}
\begin{eqnarray}
A_y(a) &=& \frac4{N_R} 
  \textrm{Im} \left\{ -\frac1{\sqrt2} U_0^{*} S_2 
  +\frac1{\sqrt3} U_1^{*}\left(\frac1{\sqrt2} S_1-S_3\right) \right.
\nonumber\\
& & -\left(\frac1{\sqrt2} S_1 +\frac12 S_3\right)^{*}\left(\frac1{\sqrt6}T_1+T_3\right) 
\nonumber\\
& &
 \left. -\frac12T_2^{*}\left(\sqrt{\frac32}T_1-T_3\right)\right\},
\\
A_y(b) &=& \frac4{N_R} \textrm{Im}\left\{\frac1{\sqrt2} U_0^{*}S_2
  -\frac1{\sqrt3}U_1^{*}\left(\frac1{\sqrt2}S_1+S_3\right) \right.
\nonumber\\
& & + \left(\frac1{\sqrt2}S_1-\frac12 S_3\right)^{*} 
    \left(\frac1{\sqrt6}T_1+T_3\right) 
\nonumber\\
& &-\frac12T_2^{*}\left. \left(\sqrt{\frac32}T_1-T_3\right)\right\}.
\end{eqnarray}
\end{subequations}

For the elastic scattering, in terms of the conventional amplitudes, Eqs.\ (\ref{eq:UaUb}), (\ref{eq:SaSb}), and (\ref{eq:TaTb}), we get
\begin{subequations}
\begin{eqnarray}
A_y(a)&=&-\frac{4\sqrt2}{N_R} 
  \textrm{Im} \left\{ U_{\alpha}^{*} \left( S_\alpha+S_\beta \right)
 +\frac14 U_{\beta}^{*} \left( -S_\alpha + S_\beta \right) \right.
\nonumber\\
&&  \left. 
  - \frac12 T_{\alpha}^{*} \left(-S_\alpha + S_\beta\right) \right\},
\\
A_y(b)&=&-\frac{4\sqrt2}{N_R} 
  \textrm{Im} \left\{U_{\alpha}^{*}\left( -S_\alpha + S_\beta \right)
  +\frac14 U_{\beta}^{*}\left( S_\alpha+S_\beta \right) \right.
\nonumber\\
&& \left. 
 - \frac12 T_{\alpha}^{*} \left( S_\alpha+S_\beta \right)\right\}, 
\end{eqnarray}
\end{subequations}
and
\begin{eqnarray}
N_R &=& 4|U_{\alpha}|^2+\frac34|U_{\beta}|^2 +
    4 \left(|S_{\alpha}|^2 +|S_{\beta}|^2\right) 
\nonumber\\
 & & +\frac12 (\tan^2\theta +4)|T_{\alpha}|^2 
\nonumber\\
 & & +2\left(\tan^2\theta +1\right) 
  \left(|T_{\beta}|^2 + \textrm{Re}(T_{\alpha}^{*} T_{\beta})\right),
\label{eq:NR_2}
\end{eqnarray}
where we used the following relation due to the time reversal relation Eq.\ (\ref{eq:T_TimeRev}), 
\begin{equation}
\textrm{Im}\left\{T_2^{*}\left(\sqrt{\frac32}T_1-T_3\right) \right\}=0.
\end{equation}

Here, we will consider the sum and the difference of $A_y(a)$ and $A_y(b)$:
\begin{subequations}
\label{eq:Aya_Ayb}
\begin{eqnarray}
&&A_y(a)+A_y(b) = -\frac{8\sqrt2}{N_R}
\nonumber\\
&\times& \textrm{Im} \left\{ \left(U_{\alpha} +\frac14U_{\beta}-\frac12 T_{\alpha}\right)^{*} S_{\beta} \right\},
\label{eq:AyaPAyb}
\\
&&A_y(a)-A_y(b)=-\frac{8\sqrt2}{N_R}
\nonumber\\
 &\times& \textrm{Im} \left\{ \left(U_{\alpha}-\frac14U_{\beta} +\frac12T_{\alpha}\right)^{*} S_{\alpha}\right\}.
\label{eq:AyaMAyb}
\end{eqnarray}
\end{subequations}
The quantities inside the curly brackets in Eqs.\ (\ref{eq:AyaPAyb}) and (\ref{eq:AyaMAyb}) are proportional to the matrix elements of $V_{\textrm{SLS}}L_1$ and $V_{\textrm{ALS}}L_1$, respectively, as shown in Eq.\ (\ref{eq:SaSb}). 
Then we can separate the contribution of the ALS interaction from that of the SLS interaction by considering such linear combinations of the analyzing powers: $A_y(a)+A_y(b)$ will be sensitive to the strength of the SLS interaction and $A_y(a)-A_y(b)$ to that of the ALS one for given $U_\alpha$, $U_\beta$ and $T_\alpha$. 
The boson-exchange model \cite{Ho89}, for instance, predicts a strong SLS interaction for the $\Sigma^{+} p$ system but a weak one for the $\Lambda p$ system. 
On the other hand, the ALS interaction is stronger for the latter than for the former, although their magnitudes are small. 
Measurements of these quantities therefore will give clear-cut examination of such characteristic features of the LS interactions.

%%%%%%%%%%%%%%%%%%%%%%%%%%%%%%%%%%%%%%%%%%%%%%%%%%%%%%%%%%%%%%
\subsection{Second order polarization observables}

First we will define the observables to be discussed. 
When the colliding particle $a$ is polarized in $i$-axis direction, one defines the depolarization $D_i^j(a)$, which describes the polarization of $a$ in $j$-axis direction after the scattering by
\begin{equation}
D_i^j(a)=\frac1{N_R} \textrm{Tr}(\bm{M} \sigma_i(a) \bm{M}^{\dag} \sigma_j(a)).
\label{eq:Depol}
\end{equation}
When we consider the polarization of the partner $b$ after the scattering, we  define the polarization transfer coefficient as
\begin{equation}
K_i^j(a \to b)=\frac1{N_R} \textrm{Tr} (\bm{M} \sigma_i(a) \bm{M}^{\dag} \sigma_j(b)).
\label{eq:SpinTrns}
\end{equation}
Finally we describe effects of the simultaneous polarizations both of $a$ and $b$ in the initial state by the spin correlation coefficient
\begin{equation}
C_{ij}=\frac1{N_R} \textrm{Tr} ( \bm{M} \sigma_i(a) \sigma_j(b) {\bm{M}}^{\dag}).
\label{eq:SpinCorr}
\end{equation}
Specifying $i$ and $j$ to two of $x$, $y$, and $z$, one gets non-vanishing fifteen observables, whose expressions for the general scattering are given in Appendix \ref{Sec:AppC}.
%which lead to the linear combinations that have characteristic features.

%%%%%%%%%%%%%%%%%%%%%%%%%%%%%%%%%%%%%%%%%%%%%%%%%%%%%%%%%%%%%%%%%%%%%%%%%%%
%%%% Linear combinations of other polarization observables in elastic scattering
%%%%%%%%%%%%%%%%%%%%%%%%%%%%%%%%%%%%%%%%%%%%%%%%%%%%%%%%%%%%%%%%%%%%%%%%%%%%%%
In the following, we will discuss linear combinations of such second order polarization observables for the elastic scattering, which are convenient for studies of characteristics of the interactions.

Let us examine the sum of the diagonal elements of the second order polarization observables.
The results are 
\begin{eqnarray}
&&D_x^x(a) + D_y^y(a) + D_z^z(a)
\nonumber \\
&=& \frac{12}{N_R} \left\{
   |U_\alpha|^2 - \frac1{16}|U_\beta|^2 \right.
\nonumber \\
& & + \frac13 \left(|S_\alpha|^2+|S_\beta|^2-4\textrm{Re}(S_\alpha^* S_\beta) \right)
\nonumber \\
& & \left.
 -\frac13 \left( |T_1|^2 + 2 |T_2|^2 + 2 |T_3|^2 \right) \right\},
\label{eq:DxxDyyDzz}
\end{eqnarray}
\begin{eqnarray}
&&K_x^x(a \to b) + K_y^y(a \to b) + K_z^z(a \to b)
\nonumber \\
&=& \frac{3}{N_R} \left\{ 2\textrm{Re}(U_\alpha^* U_\beta) 
   + \frac12 |U_\beta|^2 \right.
\nonumber \\
& & - \frac43 \left(|S_\alpha|^2-|S_\beta|^2 \right)
\nonumber \\
& & \left.
 -\frac13 \left( |T_1|^2 + 2 |T_2|^2 + 2 |T_3|^2 \right) \right\},
\label{eq:KxxKyyKzz}
\end{eqnarray}
\begin{eqnarray}
&&C_{xx} + C_{yy} + C_{zz}
\nonumber \\
&=& \frac{3}{N_R} \left\{ -|U_0|^2 + \frac13 |U_1|^2 \right.
\nonumber \\
& & - \frac43 \left(|S_\alpha|^2-|S_\beta|^2 \right)
\nonumber \\
& & \left.
 +\frac13 \left( |T_1|^2 + 2 |T_2|^2 + 2 |T_3|^2 \right) \right\},
\label{eq:CxxCyyCzz}
\end{eqnarray}
where the cross terms of the amplitudes with different ranks such as $\textrm{Re} \left( U_\beta^{*} T_\beta \right)$ are canceled out.

%which will be sensitive to the scalar amplitudes when the magnitudes of $S$'s and $T$'s are small. 

If we neglect the terms without the scalar amplitudes, for example in Eqs. (\ref{eq:DxxDyyDzz}) and (\ref{eq:NR_2}), assuming that the scalar amplitudes are dominant over the other amplitudes, we get 
\begin{eqnarray}
D &\equiv& \frac13 \left( D_x^x(a) + D_y^y(a) + D_z^z(a) \right)
\nonumber \\
&\sim& \frac1{N_R} \left( 4 |U_\alpha|^2 - \frac1{4}|U_\beta|^2 \right)
\label{eq:Dxxyyzz}
\end{eqnarray}
and
\begin{equation}
N_R \sim 4 |U_\alpha|^2 + \frac34 |U_\beta|^2,
\end{equation}
which give the magnitudes of the scalar amplitudes as
\begin{subequations}
\label{eq:D_UaUb}
\begin{eqnarray}
|U_\alpha|^2 &\sim&  \frac{N_R}{16}\left( 1 + 3D\right),
\\
|U_\beta|^2 &\sim&  {N_R}\left( 1 - D\right).
\end{eqnarray}
\end{subequations}

%Other examples for combinations of the diagonal elements of the second order polarization observables are

Moreover, for the second order polarization observables, there are several linear combinations that exhibit effects of particular components of the interaction: for example,
\begin{eqnarray}
& & C_{xx} + C_{zz} + K_x^x(a \to b) + K_z^z(a \to b)
\nonumber\\
&=&\frac8{N_R} \textrm{Re}\left\{
  \left(U_{\beta} +T_{\alpha}\right)^{*}U_{\alpha}\right\},
\end{eqnarray}
\begin{equation}
C_{xz} -C_{zx}=\frac{4\sqrt2}{N_R} \textrm{Re}\left\{
  \left( U_{\beta} +T_{\alpha} \right)^{*} S_{\alpha} \right\},
\label{eq:CxzMCzx}
\end{equation}
\begin{equation}
K_x^z(a \to b)-K_z^x(a \to b)=\frac{4\sqrt2}{N_R} \textrm{Re}\left\{
 \left( U_{\beta} +T_{\alpha} \right)^{*} S_{\beta}  \right\},
\label{eq:KxzMKzx}
\end{equation}
and
\begin{eqnarray}
D_x^z(a) +D_z^x(a) &=& -\tan\theta \left(D_x^x(a)-D_z^z(a) \right) 
\nonumber\\
&=& \frac{4\tan\theta}{N_R} \textrm{Re}\left\{
     \left(U_{\beta} +T_{\alpha}\right)^{*} 
       \left(\frac12 T_{\alpha} +T_{\beta}\right)\right\}. 
\nonumber\\
 ~
\label{eq:DxzPDzx}
\end{eqnarray}
The numerators of the right-hand sides of these equations are respectively proportional to the magnitudes of the amplitudes, $U_{\alpha}$, $S_{\alpha}$, $S_{\beta}$, and $\frac12 T_{\alpha}+T_{\beta}$, and then the left-hand side quantities will give some kinds of measures of the strength of the corresponding interactions, when multiplied by the cross section.
Since Eqs.\ (\ref{eq:CxzMCzx}), (\ref{eq:KxzMKzx}), and (\ref{eq:DxzPDzx}) include only $U_{\beta}$ as the scalar amplitude, they will be useful for investigations of the spin-spin interaction.

%%%%%%%%%%%%%%%%%%%%%%%%%%%%%%%%%%%%%%%%%%%%%
%%%% Spin-dependent Total Cross Sections %%%%
%%%%%%%%%%%%%%%%%%%%%%%%%%%%%%%%%%%%%%%%%%%%%
\subsection{Spin-dependent total cross sections}
\label{subsec:SPTCS}

Total cross sections, when no Coulomb interaction acts, provide the imaginary parts of forward scattering amplitudes by the optical theorem. 
In the present system, three amplitudes with $\kappa=0$, namely $U_{\alpha}$, $U_{\beta}$, and $T_1$, survive at the forward angle, which are important sources of information on the scalar and tensor interactions. 
We will consider correspondingly three kinds of total cross sections by choosing proper polarizations of target and beam particles. 

Let us denote the spin density of an initial state, which consists of the beam particle $a$ and the target particle $b$, by $\rho_{\textrm{i}}^{(a,b)}$.
Then the optical theorem gives the corresponding total cross section $\sigma$ as
\begin{equation}
\sigma = 
 \frac{4\pi}k \textrm{Im} \left\{ \textrm{Tr}
  \left( \rho_{\textrm{i}}^{(a,b)} \bm{M} \right)_{\theta=0} \right\},
\label{eq:Opt_Th}
\end{equation}
where $k$ is the magnitude of $\bm{k}_{\textrm{i}}$.

One of the independent total cross sections is the unpolarized total cross section $\sigma_{\textrm{unpol}}$ calculated from a density matrix 
\begin{equation}
\rho_{\textrm{i}}^{(a,b)} = \frac12 I^{(a)} \otimes \frac12 I^{(b)},
\end{equation}
where $I^{(a)}$ and $I^{(b)}$ are the unit matrices for the particles $a$ and $b$, respectively.

As for the other two total cross sections, we consider those with longitudinal and transverse polarizations. 
In general, when the particles $a$ and $b$ are polarized in the $j$-axis direction with polarizations $p^{(a)}$ and $p^{(b)}$, the corresponding cross section $\sigma_j (p^{(a)}, p^{(b)})$ is obtained with spin density matrix 
\begin{equation} 
\rho_{\textrm{i}}^{(a,b)} = 
  p^{(a)} \frac{1}2 \sigma_j^{(a)} \otimes p^{(b)} \frac{1}2 \sigma_j^{(b)}, 
\end{equation} 
where $\bm{\sigma}^{(a)}$ and $\bm{\sigma}^{(b)}$ are the Pauli spin matrices for $a$ and $b$. 
The polarization axis is along the beam direction, $j=z$, for the longitudinal configuration and is perpendicular to the beam direction, typically $j=y$, for the transverse one. 
The longitudinal asymmetry $\Delta\sigma_{\textrm{L}}$ and the transverse asymmetry $\Delta\sigma_{\textrm{T}}$ \cite{Is01} are defined as the difference of the cross sections provided by the reversal of the particle $b$'s spin:
\begin{subequations}
\begin{eqnarray}
\Delta\sigma_{\textrm{L}} &=& \sigma_z\left(+1,-1\right) - \sigma_z\left(+1,+1\right), 
\\
\Delta\sigma_{\textrm{T}} &=& \sigma_y\left(+1,-1\right) - \sigma_y\left(+1,+1\right).
\end{eqnarray}
\end{subequations}

From these definitions, we obtain
s%
\begin{subequations}
\begin{eqnarray}
\textrm{Im} \left( U_{\alpha}(\theta=0) \right) &=& 
  \frac{k}{4\pi} \sigma_{\textrm{unpol}},
\\
\textrm{Im} \left(U_{\beta}(\theta=0)\right) 
  &=& - \frac{k}{6\pi} \left( \Delta\sigma_{\textrm{L}} 
  + 2 \Delta\sigma_{\textrm{T}} \right),
\\
\textrm{Im} \left( T_1(\theta=0) \right) 
 &=&  - \frac{{\sqrt2}k}{4{\sqrt3}\pi} \left( \Delta\sigma_{\textrm{L}}
   - \Delta\sigma_{\textrm{T}} \right).
\end{eqnarray}
\end{subequations}
That is, for the imaginary part of the forward scattering amplitude, the spin-independent scalar amplitude is determined by $\sigma_{\textrm{unpol}}$, and the spin-spin scalar amplitude and the tensor amplitude are determined by $\Delta\sigma_{\textrm{L}}$ and $\Delta\sigma_{\textrm{T}}$. 
Then measurements of such total cross sections will provide criticisms of the calculated amplitudes for scattering of neutral hyperons, for example, $\Lambda p$ scattering. 
Since the imaginary parts of scattering amplitudes reflect absorption effects due to related reaction channels, measurements of these cross sections will provide information of nature of the couplings with the channels, particularly, by clarifying which kinds of the spin-dependent interactions are important.

%%%%%%%%%%%%%%%%%%%%%%%%%%%%%%%%%%%%%%%%%%%%%%%%%%%%%%%%%%%%%%
\section{Numerical examination in $\Sigma^{+} p$ scattering}
\label{Sec:IV}
%%%%%%%%%%%%%%%%%%%%%%%%%%%%%%%%%%%%%%%%%%%%%%%%%%%%%%%%%%%%%%

%%%% Figure 1 %%%%
\begin{figure}[t]
\includegraphics[scale=0.45]{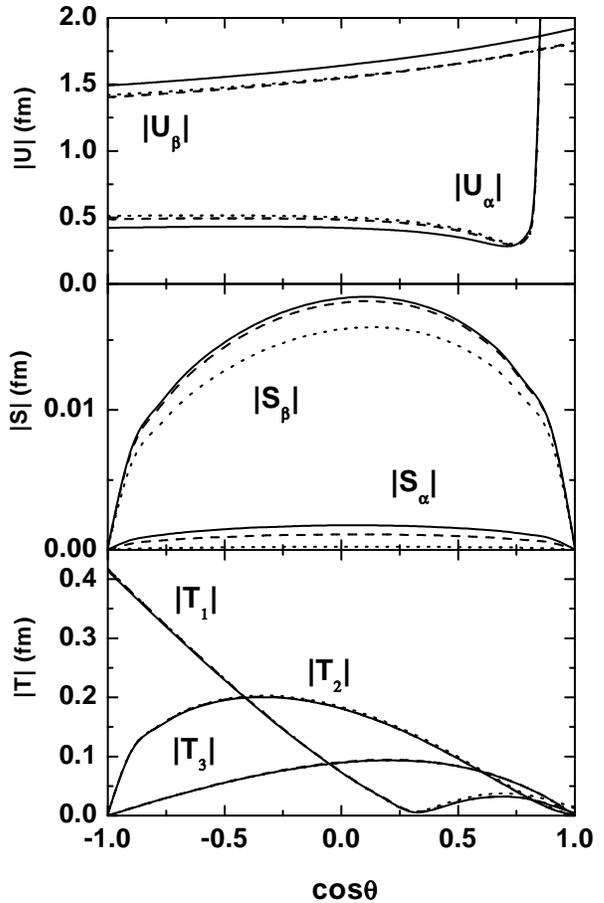}
%\vspace{3cm}
\caption{
The magnitudes of the amplitudes, $U_{\alpha}$, $U_{\beta}$, $S_\alpha$, $S_\beta$, $T_1$, $T_2$, and $T_3$ for the $\Sigma^{+} p$ scattering at $p_{\Sigma^{+}}=170$ MeV/c.
The solid lines are calculations by the NSC97a  potential model, the dashed lines by the NSC97c one, and the dotted lines by the NSC97f one. 
\label{ia170}
}
\end{figure}

%%%% Figure 2 %%%%
\begin{figure*}[t]
\includegraphics[scale=0.5]{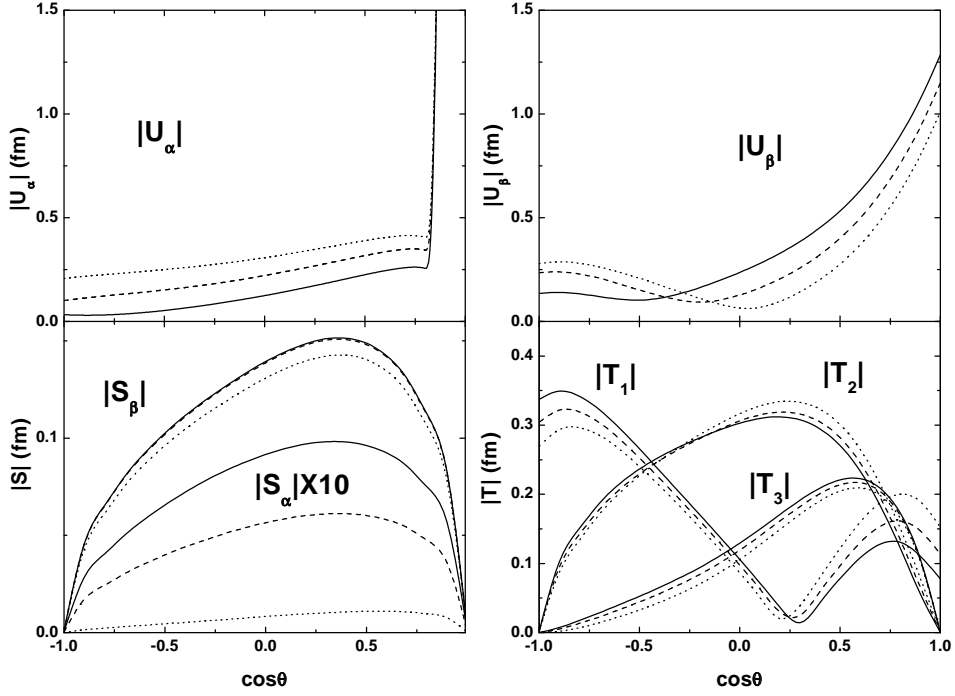}
%\vspace{3cm}
\caption{
The same as Fig. \protect\ref{ia170}, but at $p_{\Sigma^{+}}=450$ MeV/c.
\label{ia450}}
\end{figure*}

As a test of the validity of the theoretical predictions, we will perform numerical calculations for $\Sigma^{+} p$ scattering with the Nijmegen soft-core One-Boson-Exchange potential models (NSC97) \cite{Ri99}. 
In Ref.\ \cite{Ri99}, six different $YN$ potential models, NSC97a to NSC97f, which are characterized by different choices for the magnetic vector ratio, have been phenomenologically derived as descriptions of existing experimental data.
These potential models contain the scalar, vector, and tensor components of various strengths, and then should be suitable for the present test.
In all of the figures below, we will plot results of three potentials, namely NSC97a, NSC97c, and NSC97f, to avoid an unnecessary confusion due to overclosed lines.

%%%%%%%%%%%%%%%%%%%%%%%%%%%%%
%%% Scattering Amplitudes %%%
%%%%%%%%%%%%%%%%%%%%%%%%%%%%%
\subsection{Scattering amplitudes}
\label{sec:SA_NSC97}

The magnitudes of the scalar amplitudes, $U_{\alpha}$ and $U_{\beta}$, the vector amplitudes, $S_\alpha$ and $S_\beta$, and the tensor amplitudes, $T_1$, $T_2$, and $T_3$, for the $\Sigma^{+} p$ scattering at $p_{\Sigma^{+}}=$ 170 and 450 MeV/c are plotted in Figs.\ \ref{ia170} and \ref{ia450}, respectively.
As seen in these figures, %Figs.\ \ref{ia170} and \ref{ia450}, 
the NSC97 models give similar angular dependence for each kind of the amplitude in a global sense: 
a weak $\cos\theta$ dependence of the scalar amplitudes except for forward angles particularly remarkable at $p_{\Sigma^{+}}=$ 170 MeV/c, a hill-like distribution peaked at $\cos\theta=0$ to 0.5 for the vector amplitudes, an one-node-like structure for the tensor amplitude $T_1$, etc.
%
% PWIA
%
Such characteristics of the angular dependence will be understood by the plane-wave Born approximation as demonstrated in Ref.\ \cite{Oh84}, where the matrix element of the coordinate-space tensor in Eq.\ (\ref{eq:M_Kkappa}) is given as
\begin{eqnarray}
&& \langle \bm{k}_{\textrm{f}} | \bm{R}^{(K)}_{\kappa} | \bm{k}_{\textrm{i}} \rangle
%\nonumber\\
%&=& \langle \bm{k}_{\textrm{f}} |Y_{K \kappa} V_K | \bm{k}_{\textrm{i}} \rangle
\nonumber \\
 &=& \int e^{i\bm{q}\cdot\bm{r}} Y_{K \kappa}(\hat{\bm{r}}) V_{K}(r) d\bm{r}
\nonumber \\
&=&  4\pi i^K Y_{K \kappa}(\hat{\bm{q}}) \int_0^\infty j_K(qr) V_K(r) r^2 dr.
\label{eq:Mq}
\end{eqnarray}
Here, $V_K$ is a relevant potential for the tensor of rank $K$, and $\bm{q}$ is the momentum transfer in the scattering,
\begin{equation}
\bm{q} = \bm{k}_{\textrm{i}} - \bm{k}_{\textrm{f}}.
\end{equation}
%and $\bm{r}$ is the variables of relative motion.
%
For the elastic scattering, the magnitude and  azimuthal angle of the momentum transfer $\bm{q}$ are given by
\begin{subequations}
\label{eq:q_mag_azim}
\begin{eqnarray}
q &=&  k \sqrt{2(1-\cos\theta)},
\\
\tan\theta_q &=& \frac{\sin\theta}{1-\cos\theta}, 
\end{eqnarray}
\end{subequations}
where $k = |\bm{k}_{\textrm{i}}| = |\bm{k}_{\textrm{f}}|$. 
Eqs.\ (\ref{eq:Mq}) and (\ref{eq:q_mag_azim}) explain essential features of the angular dependence of the amplitudes in Figs.\ \ref{ia170} and \ref{ia450}.

Eq.\ (\ref{eq:Mq}) indicates that the magnitudes of the amplitudes in Figs.\ \ref{ia170} and \ref{ia450} are measures of the strengths of the relevant interactions.
At $p_{\Sigma^{+}}=170$ MeV/c, the magnitude of $U_{\beta}$ is much larger than that of $U_{\alpha}$  and of the vector and tensor amplitudes except for $\cos\theta \sim 1$ where the Coulomb scattering is dominant in $U_{\alpha}$. 
Such superiority of the spin-spin amplitudes is interpreted as the result of large contributions of the pion-exchange mechanism to the central interaction.
On the other hand, at $p_{\Sigma^{+}}=450$ MeV/c, the magnitudes of $U_{\alpha}$, $U_{\beta}$, and the tensor amplitudes are comparable.
The magnitude of $S_\alpha$ is very small, indicating weak ALS interactions. 
Both of $|S_\alpha|$ and $|S_\beta|$ decrease with the change of the interaction from the NSC97a to the NSC97f, reflecting stronger LS interactions in the NSC97a  and weaker ones in the NSC97f.

%%%%%%%%%%%%%%%%%%%%%%
%%% Cross Sections %%%
%%%%%%%%%%%%%%%%%%%%%%
\subsection{Cross sections}

%%%% Figure 3 %%%%
\begin{figure*}[t]
\includegraphics[scale=0.4]{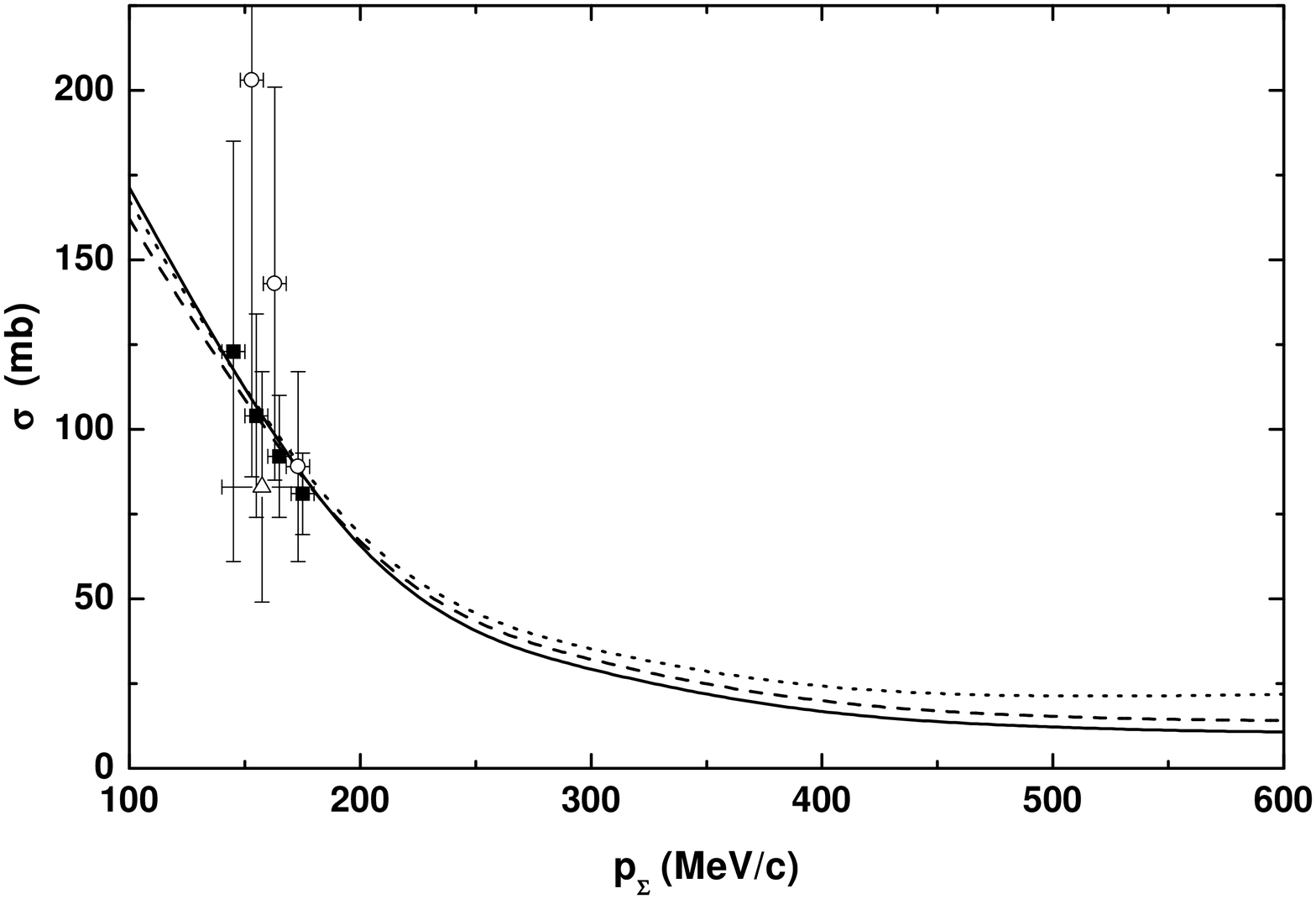}
%\vspace{3cm}
\caption{
"Total" cross section defined by Eq.\ (\protect\ref{eq:qttl}) for the $\Sigma^{+} p$ scattering for 100 MeV/c $\le p_{\Sigma^{+}} \le$ 600 MeV/c.
See the caption of Fig.\ \protect\ref{ia170} for the definitions of the theoretical curves. 
The filled squares denote the experimental data from \protect\cite{Ei71}, the open circles from Ref.\ \protect\cite{Do66}, the open triangle from \protect\cite{Ru67}.
\label{spttl}}
\end{figure*}

%%%% Figure 4 %%%%
\begin{figure}[t]
\includegraphics[scale=0.45]{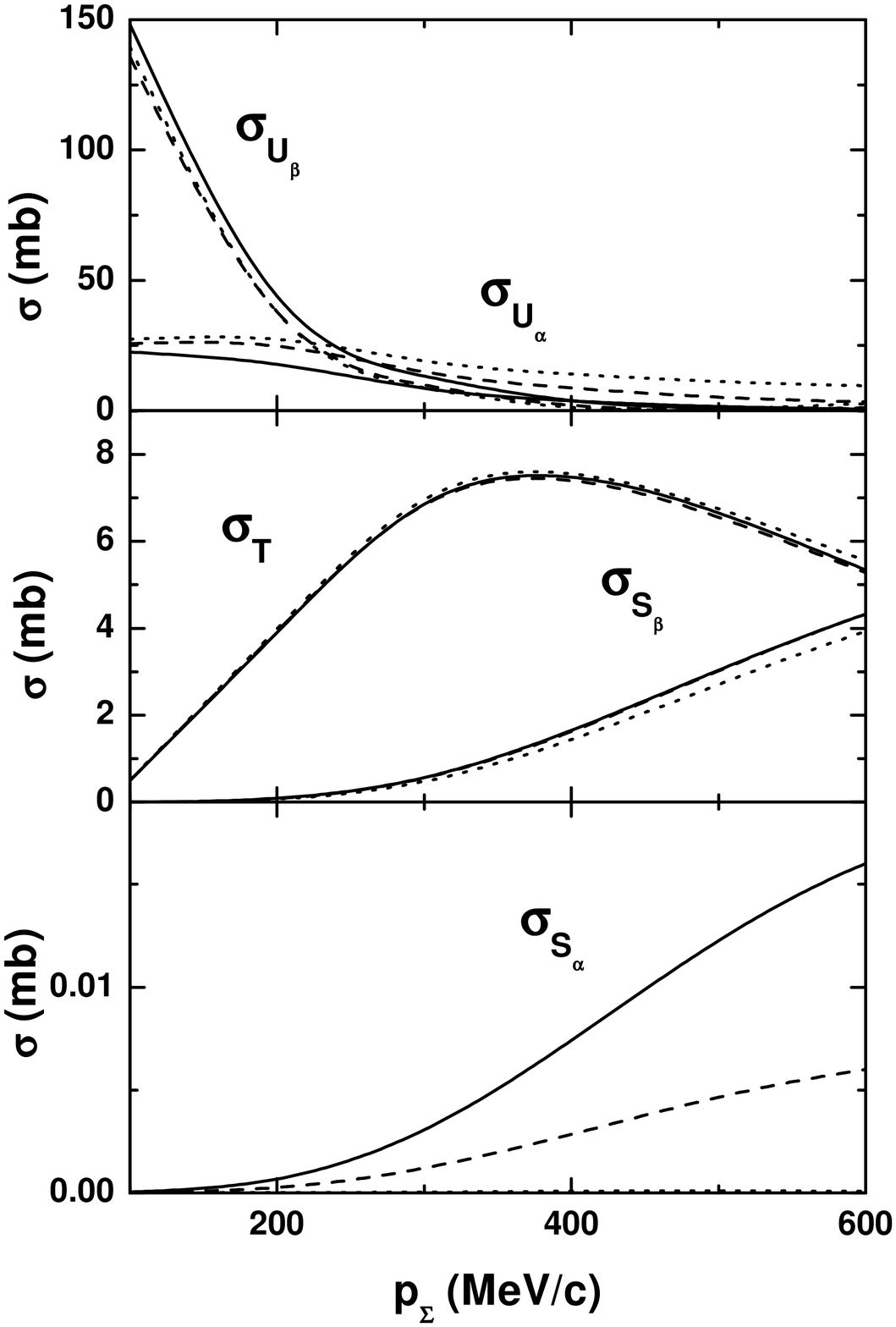}
%\vspace{3cm}
\caption{
Decomposition of total cross section of the $\Sigma^{+} p$ scattering into spin-space components defined in Eq.\ (\protect\ref{eq:DCS_deomp}). 
See the caption of Fig.\ \protect\ref{ia170} for the definitions of the theoretical curves.
\label{ttlcomp}}
\end{figure}

%%%%%%%%%%%%%%%%%%%%%%%%%%%%%%%
%%%% "Total" cross section %%%%
%%%%%%%%%%%%%%%%%%%%%%%%%%%%%%%

As discussed in Subsec.\ \ref{subsec:SPTCS}, a set of the spin-dependent total cross sections for a system without a Coulomb force works as measures of the strength for the spin-dependent interactions by the optical theorem. 
However, the Coulomb interaction in the $\Sigma^{+}p$ system prevents such a measurement of the total cross section. 
In analyzing $YN$ data of 1960's, an averaged value of the cross section over a certain range of the scattering angle, $\cos\theta_{\textrm{min}}$ to $\cos\theta_{\textrm{max}}$, 
\begin{equation}
\sigma = \frac{2}{\cos\theta_{\textrm{max}}-\cos\theta_{\textrm{min}}} 
\int_{\cos\theta_{\textrm{min}}}^{\cos\theta_{\textrm{max}}} \frac{d\sigma(\theta)}{d\cos\theta} d\cos\theta 
\label{eq:qttl}
\end{equation}
was used as "total" $\Sigma^{+}p$ cross sections. 
%Parameters of the NSC97 models are fitted to such experimental $YN$ data \cite{Ei71}.
In Fig.\ \ref{spttl}, the "total" cross section with $\cos\theta_{\textrm{min}}=-0.5$ and $\cos\theta_{\textrm{max}}=0.5$ \cite{Ri99} calculated for the NSC97a, NSC97c, and NSC97f is compared with the experimental data \cite{Ei71,Do66,Ru67}.
All of the measured cross sections are localized in the low momentum region below $p_{\Sigma^{+}} = 200$ MeV/c, where the cross section calculated by any version of the interaction has a similar magnitude and agrees to such low momentum data. 

As indicated in Eqs.\ (\ref{eq:Difcs}) and (\ref{eq:NR_2}), the differential cross section consists of the absolute squares of scalar, vector, and tensor amplitudes.
The "total" cross section accordingly consists of the corresponding contributions:
\begin{equation}
\sigma_{A} = \frac{2\pi k_{\textrm{f}}}{4k_{\textrm{i}}}
  \frac{2}{\cos\theta_{\textrm{max}}-\cos\theta_{\textrm{min}}} 
\int_{\cos\theta_{\textrm{min}}}^{\cos\theta_{\textrm{max}}}
O_A d\cos\theta,
\label{eq:DCS_deomp}
\end{equation}
where $O_A = 4|U_{\alpha}|^2, \frac34|U_{\beta}|^2, 4 |S_{\alpha}|^2, 4 |S_{\beta}|^2$, and $|T_1|^2 + 2 \left( |T_2|^2 +|T_3|^2 \right)$ for 
$A=U_{\alpha}, U_{\beta}, S_{\alpha}, S_{\beta}$, and $T$, respectively.
Such components of the cross sections are displayed in Fig.\ \ref{ttlcomp}. 
It is seen that the "total" cross section is mainly governed by the contributions of the scalar amplitudes and the contribution of the spin-spin interaction, $U_{\beta}$, is particularly dominant at the low momenta. 
This $U_\beta$ contribution explains the main part of the measured cross sections in Fig. \ref{spttl}.
For higher momentum region, the contribution from the tensor amplitudes becomes comparable with that from the scalar amplitudes as indicated in Fig.\ \ref{ia450}.

%%%%%%%%%%%%%%%%%%%%%%%%%%%%%
%%% Observables (dcs, Ay) %%%
%%%%%%%%%%%%%%%%%%%%%%%%%%%%%

\begin{figure}[t]
\includegraphics[scale=0.45]{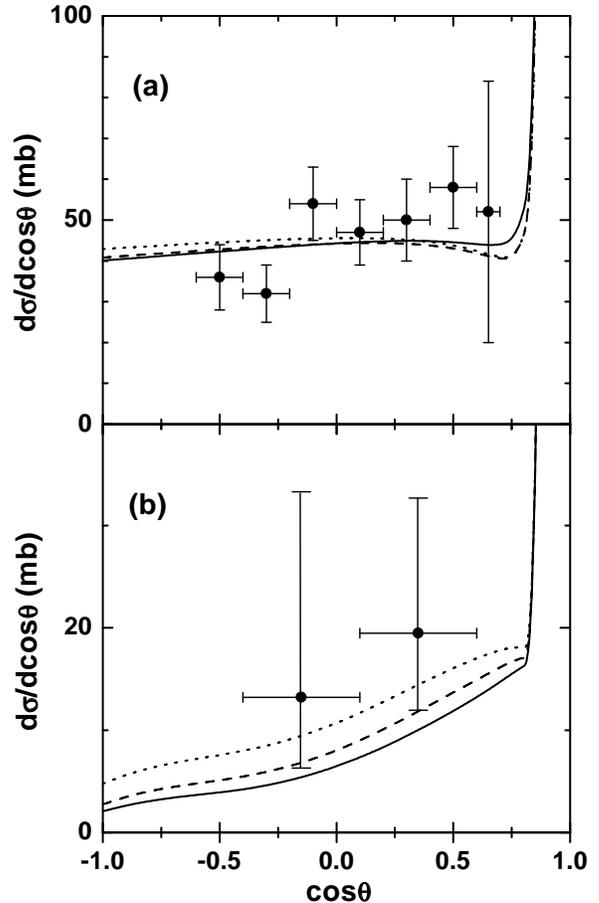}
%\vspace{3cm}
\caption{
Differential cross section $d\sigma(\theta)/d\cos\theta$ for the $\Sigma^+ p$ scattering at $p_{\Sigma^{+}}=$ 170 MeV/c (a) and 450 MeV/c (b). 
See the caption of Fig.\ \protect\ref{ia170} for the definitions of the theoretical curves.
Experimental data are taken from Refs.\ \protect\cite{Ei71} for $p_{\Sigma^{+}}=$ 170 MeV/c and from Ref.\ \protect\cite{Ah99} for $p_{\Sigma^{+}}=$ 450 MeV/c.
\label{sp-dcs}
}
\end{figure}

%%%%%%%%%%%%%%%%%%%%%%%%%%%%%%%%%%%%%%%%%%%%%%%%%%%%%%%%%%%
In Fig.\ \ref{sp-dcs}, the calculated differential cross sections $d\sigma(\theta)/d\cos\theta$ at $p_{\Sigma^{+}}=170$ and $450$ MeV/c 
%by the NSC97a, NSC97c, and NSC97f models 
are displayed, which are very similar to each other for given $p_{\Sigma^{+}}$ and agree with the experimental data \cite{Ei71,Ah99}. 
Since  $|T_j|^2$ $(j=1, 2, 3)$ and $|S_j|^2$ $(j=\alpha, \beta)$ are small compared to $|U_{\alpha}|^2$ and $|U_{\beta}|^2$ for $p_{\Sigma^{+}}=170$ MeV as shown in Fig.\ \ref{ia170}, the differential cross section is governed mainly by $4|U_{\alpha}|^2+\frac34|U_{\beta}|^2$ according to Eq.\ (\ref{eq:NR_2}), where the calculated $|U_{\alpha}|$ and $|U_{\beta}|$ complement with each other:  $|U_{\beta}|$ by the NSC97a potential is larger than that by the other versions of the interaction as shown in Fig.\ \ref{ia170} but the excess is compensated by the smallness of $|U_{\alpha}|$, giving the resultant cross section similar to other calculations' as seen Fig.\ \ref{sp-dcs} (a). 

%for instance at $p_{\Sigma^+}$ = 170 MeV/c,

%%%% Figure 5 %%%%

%%%%%%%%%%%%%%%%%%%%%%%%%%%%%%%
\subsection{Analyzing powers}
%%%%%%%%%%%%%%%%%%%%%%%%%%%%%

%%%% Figure 6 %%%%
\begin{figure}[t]
\includegraphics[scale=0.45]{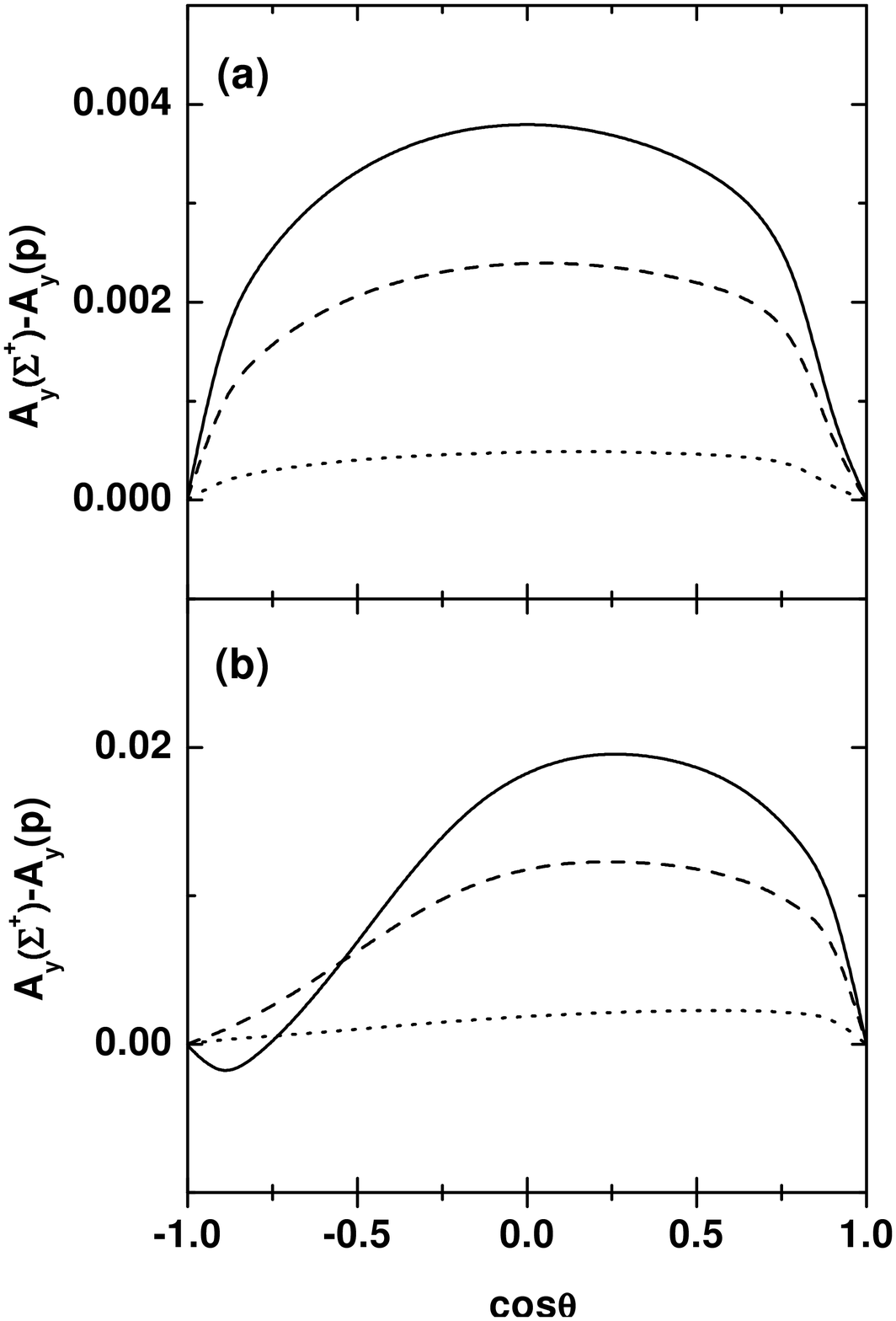}
%\vspace{3cm}
\caption{
The difference of the vector analyzing powers $A_y(\Sigma^{+})-A_y(p)$ for the $\Sigma^+ p$ scattering at $p_{\Sigma^{+}}=$ 170 MeV/c (a) and 450 MeV/c (b). 
See the caption of Fig.\ \protect\ref{ia170} for the definitions of the theoretical curves.
\label{sp-aydif}
}
\end{figure}

%%%% Figure 7 %%%%
\begin{figure}[t]
\includegraphics[scale=0.45]{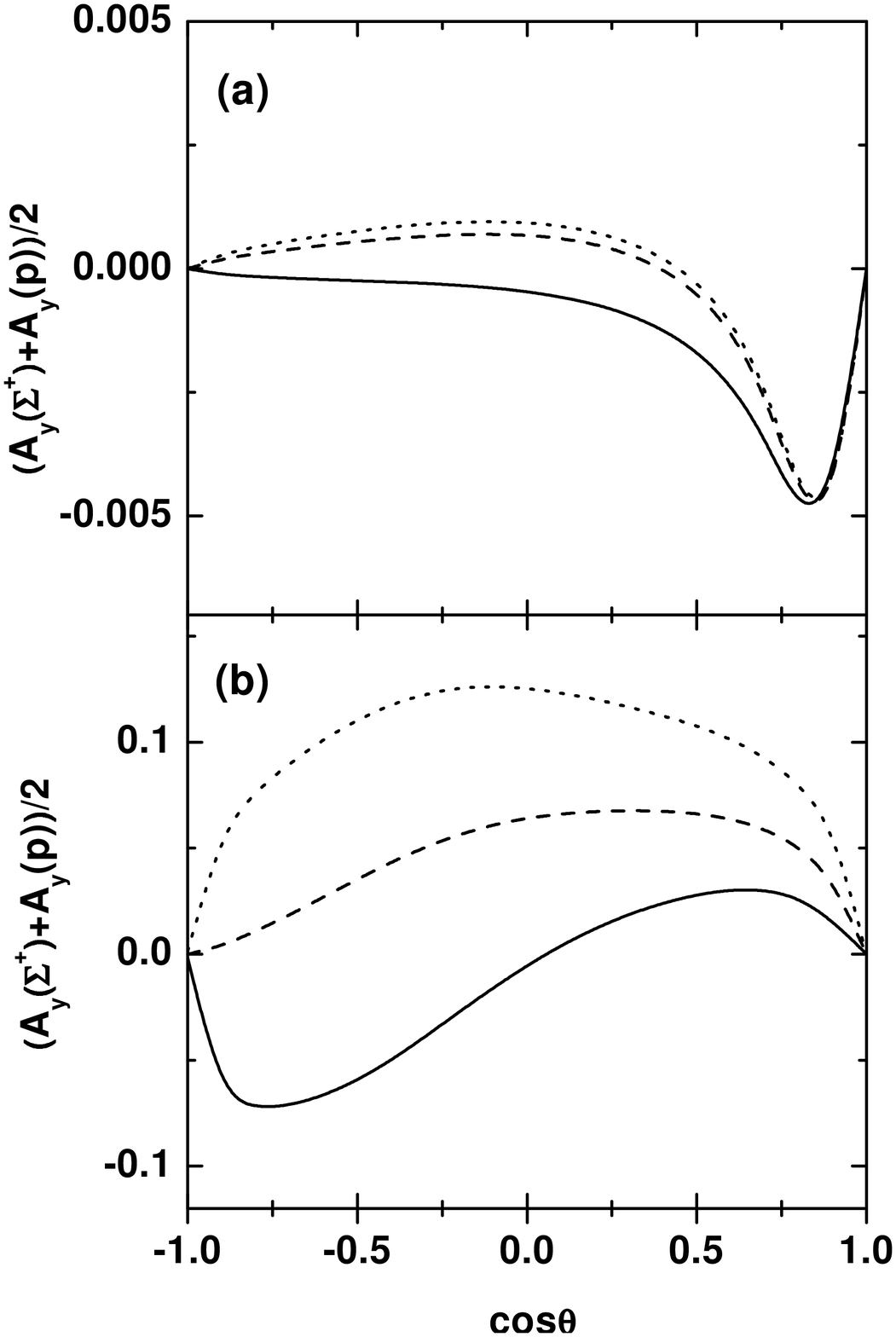}
%\vspace{3cm}
\caption{
The average of the vector analyzing powers $\left( A_y(\Sigma^{+})+A_y(p)\right)/2$ for the $\Sigma^+ p$ scattering at $p_{\Sigma^{+}}=$ 170 MeV/c (a) and 450 MeV/c (b). 
See the caption of Fig.\ \protect\ref{ia170} for the definitions of the theoretical curves.
\label{sp-ayavr}
}
\end{figure}

The calculated difference and average of the $\Sigma^{+}$ analyzing power $A_y(\Sigma^{+})$ and the proton analyzing power $A_y(p)$, 
\begin{subequations}
\label{eq:Ay_avr_diff}
\begin{eqnarray}
\Delta A_y &=& A_y(\Sigma^{+}) - A_y(p),
\\
A_y^{\textrm{avr}} &=& \frac12 \left( A_y(\Sigma^{+}) + A_y(p) \right)
\end{eqnarray}
\end{subequations}
are plotted in Figs.\ \ref{sp-aydif} and \ref{sp-ayavr} for the $\Sigma^+ p$ scattering at $p_{\Sigma^{+}}=$ 170 and 450 MeV/c. 
Contrary to the similarity in the calculated differential cross sections among the NSC97 models, the difference in the linear combinations of the calculated vector analyzing powers among the NSC97 models is significant. 
The calculated difference  $\Delta A_y$  reflects evenly the magnitude of the ALS interaction, giving the largest magnitude for the NSC97a model and the smallest one for the NSC97f model. 
On the other hand, the average of the analyzing powers $A_y^{\textrm{avr}}$ is rather confusing. 
While the NSC97a and the NSC97c give almost the same magnitudes of the SLS amplitude, which are larger than that of the NSC97f, as shown in Figs.\ \ref{ia170} and \ref{ia450}, the calculations of $A_y^{\textrm{avr}}$ do not reflect this tendency.   
Particularly $A_y^{\textrm{avr}}$ for the NSC97a potential has the opposite sign to the one for other versions at some angles. 
This happens due to the sensitivity of $A_y^{\textrm{avr}}$ not only on the SLS amplitude $S_\alpha$ but also on a combination of the scalar amplitudes, the spin independent amplitude $U_\alpha$ and the spin-spin one $U_\beta$ as seen Eq.\ (\ref{eq:AyaPAyb}). 
The dependence on the combination of scalar amplitudes overrides that on the SLS amplitude in $A_y^{\textrm{avr}}$, while this is not the case for $\Delta A_y$.
We therefore conclude that the spin-independent and spin-spin central interactions in $YN$ scattering should be determined from other sources in order to obtain unique information of LS interactions from measurements of analyzing powers.

%%%%%%%%%%%%%%%%%%%%%%%%%%%%%
\subsection{Depolarizations}
%%%%%%%%%%%%%%%%%%%%%%%%%%%%

\begin{figure}[t]
\includegraphics[scale=0.45]{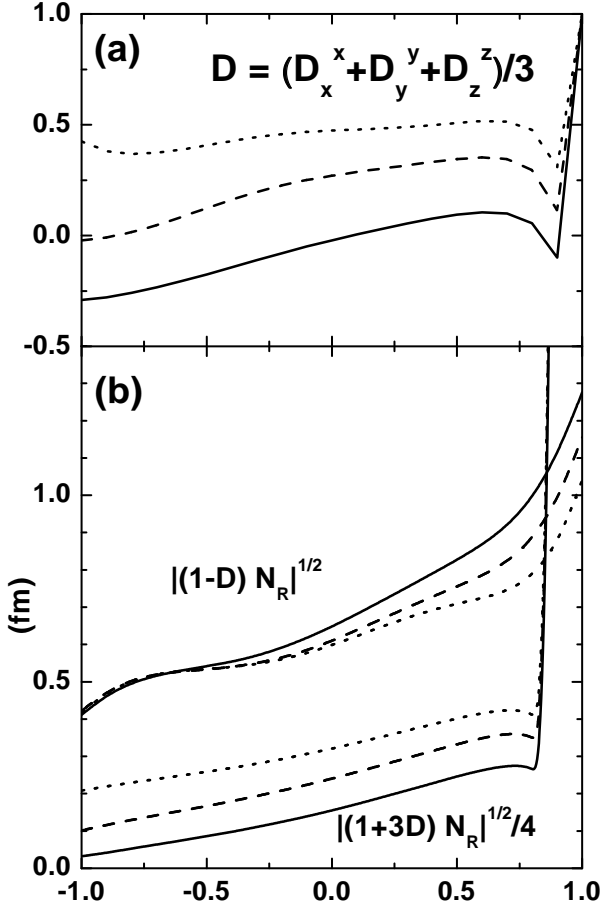}
%\vspace{3cm}
\caption{
(a) The average of the diagonal depolarizations of $\Sigma^{+}$ by Eq.\ (\protect\ref{eq:Dxxyyzz}), and (b) the extracted scalar amplitudes by Eq.\ (\protect\ref{eq:D_UaUb})  for the $\Sigma^+ p$ scattering at $p_{\Sigma^{+}}=$  450 MeV/c. 
See the caption of Fig.\ \protect\ref{ia170} for the definitions of the theoretical curves.
\label{dxxyyzz}
}
\end{figure}

Information of the scalar interactions can be obtained from, for example, depolarizations as discussed in the preceding section. 
Fig.\ \ref{dxxyyzz} displays the average of the diagonal elements of depolarizations, $D$ in Eq.\ (\ref{eq:Dxxyyzz}), for the $\Sigma^+ p$ scattering at $p_{\Sigma^{+}}=$ 450 MeV/c and the quantities defined as
\begin{subequations}
\begin{eqnarray}
|\tilde{U}_{\alpha}| &\equiv& \left|N_R(1+3D)\right|^{1/2}/4,
\\
|\tilde{U}_{\beta}| &\equiv& \left|N_R(1-D)\right|^{1/2},
\end{eqnarray}
\end{subequations}
which are predicted  to give scalar amplitudes $|U_{\alpha}|$ and $|U_{\beta}|$, respectively, by Eq.\ (\ref{eq:D_UaUb}).
In the figure, the average depolarization $D$ discriminates between the three versions of the NSC97 interaction, particularly with different signs at backward angles for the NSC97a and NSC97f.
The extracted scalar amplitudes $|\tilde{U}_{\alpha}|$ and $|\tilde{U}_{\beta}|$ follow the tendency of the amplitudes in Fig.\ \ref{ia450} except $|\tilde{U}_{\beta}|$ at backward angles, where effects from the tensor amplitude $T_1$ may not be neglected. 
Thus $\sigma(1+3D)$ and $\sigma(1-D)$ at middle and forward angles will be good measures of the spin-independent and spin-spin central interactions. 
It should be noted that the components of $D$, i.e., $D_x^x$, $D_y^y$, and $D_z^z$, also discriminate the above versions of the interaction as well as $D$, although their interaction dependence is not displayed at present.
However, $D$ is favorable to identify the contribution of the spin-independent interaction and that of the spin-spin one separately.

%%%% Figure 8 %%%%

%%%%%%%%%%%%%%%%%
\section{Summary}
\label{Sec:V}

In the $YN$ scattering, we have investigated the contributions of the spin-dependent interactions to the observables by decomposing the scattering amplitudes according to the tensorial property in the spin space so that the contributions of the interactions are individually identified. 
In terms of such amplitudes, the expressions of the polarization observables are derived for general scattering.  

For the elastic scattering, we have found some linear combinations of the observables to be sensitive to  particular interactions and thus to be  favorable for studies of contributions of the interactions. 
In fact, the contributions of the SLS interaction and those of the ALS one are separated to each other by considering the linear combinations of the vector analyzing powers, the spin correlation coefficients, the polarization transfer coefficients, etc., each of which is proportional to the strength of the SLS or ALS interactions. 
Similar linear combinations have been found to be sensitive to the tensor interactions and the spin-independent and spin-spin central ones.

A part of the theoretical predictions is numerically examined for the $\Sigma^{+}p$ scattering as an example. 
The observables have been calculated by the use of the series of the NSC97 interactions and it has been found that some linear combinations of the observables are useful to discriminate the different versions of the interaction even when their cross sections are similar to be indistinguishable. 
 
The total cross section has been investigated for the unpolarized beam and target as well as for the longitudinal-polarized ones and for the transverse-polarized ones. 
These provide the imaginary parts of the amplitude of the forward scattering by the spin-independent central interactions, the spin-spin central ones, and the tensor ones. 
Measurements of these cross sections in scattering of neutral hyperons by the  proton therefore will provide important information of the interactions, particularly of the nature of the coupling with the related reaction channels. 

Due to the significance of information on the $YN$ interaction, we hope more experiments to be performed for polarization phenomena in the $YN$ scattering so that the details of the interactions, which include the spin dependence, will be determined.

%%%%%%%%%%%%%%%%%%%%%%%%%%%%%%%%%%%%%%%%%%%%%%%%%%%%%%%%%%%%%%%%%%%%%%%%%%%
\begin{acknowledgments}
The authors would like to express their thanks to Dr. E. Hiyama for stimulating them by providing a lot of information on hypernucleus structures. 
This research was supported by the Japan Society for the Promotion of Science, 
under a Grant-in-Aid for Scientific Research (Grant No. 13640300).
\end{acknowledgments}

%%%%%%%%%%%
\appendix

%%%%%%%%%%%%%%%%%%%%%%%%%%%%%%%%%%%%%%%%%%%%%%%%%%%%%%%%%%%%%%%%%
\section{Spin-space tensor components of scattering amplitude}
\label{Sec:AppA}

The T-matrix $\bm{M}$ for general scattering $a + b \to c+ d$ with a given parity, is described as
\begin{equation}
\bm{M}=\left(
\begin{array}{cccc}
 A & B & C &  D \\
 E & F & G &  H \\
-H & G & F & -E \\
 D &-C &-B &  A
\end{array}
\right),
\label{eq:M-elements}
\end{equation}
where the row is designated by the spin $z$ components $\nu_a$ and $\nu_b$ as $(\nu_a,\nu_b)=(1/2, 1/2)$, $(1/2, -1/2)$, $(-1/2, 1/2)$, and $(-1/2, -1/2)$ from left to right, and the column by $\nu_c$ and $\nu_d$ as $(\nu_c,\nu_d)=(1/2, 1/2)$, $(1/2, -1/2)$, $(-1/2, 1/2)$, and $(-1/2, -1/2)$ from top to bottom. 
Applying Eq.\ (\ref{eq:Melemnts}) to $A$, \dots, $H$ in Eq.\ (\ref{eq:M-elements}), we get    
\begin{subequations}
\label{eq:AH_UST}
\begin{eqnarray}
A&=&\frac1{\sqrt3} U_1 +\frac1{\sqrt6} T_1,
\\
B&=&\frac1{\sqrt2} S_1 +\frac12 S_3 -\frac12 T_2,
\\
C&=&-\frac1{\sqrt2} S_1 +\frac12 S_3 -\frac12 T_2,
\\
D&=&T_3,
\\
E&=&\frac1{\sqrt2} S_2 -\frac12 S_3 -\frac12 T_2, 
\\
F&=&\frac12 U_0 +\frac1{2\sqrt3} U_1 -\frac1{\sqrt6} T_1,
\\
G&=&-\frac12 U_0 +\frac1{2\sqrt3} U_1 -\frac1{\sqrt6} T_1, 
\\
H&=&\frac1{\sqrt2}S_2 +\frac12 S_3 +\frac12 T_2.
\end{eqnarray}
\end{subequations}

Conversely one can calculate $U_j$, $S_j$, and $T_j$ from $A$, \dots, $H$ 
\begin{subequations}
\label{eq:UST_AH}
\begin{eqnarray}
U_0 &=& F-G,
\\
U_1 &=& \frac1{\sqrt3}(2A+F+G),
\\
S_1 &=& \frac1{\sqrt2}(B-C),
\\
S_2 &=& \frac1{\sqrt2}(E+H),
\\
S_3 &=& \frac12(B+C-E+H),
\\
T_1 &=& \sqrt{\frac23}(A-F-G), 
\\
T_2 &=& -\frac12 (B+C+E-H), 
\\
T_3 &=& D.
\end{eqnarray}
\end{subequations}

%%%%%%%%%%%%%%%%%%%%%%%%%%%%%%%%%%%%%%%%%%%%%%%%%%%%%%%%%%%%%%%%%
\section{Time reversal theorem in elastic scattering}
\label{Sec:AppB}

In this appendix, we will give the derivation of the relationships due to the time reversal theorem for the vector amplitudes Eq.\ (\ref{eq:S_TimeRev}) and the tensor ones Eq.\ (\ref{eq:T_TimeRev}).

The time reversal theorem is described as \cite{Go64}
\begin{eqnarray}
&&\langle\nu_c  \nu_d; \bm{k}_{\textrm{f}} | \bm{M} |\nu_a  \nu_b; \bm{k}_{\textrm{i}} \rangle =  (-)^{\nu_c+\nu_d-\nu_a-\nu_b} 
\nonumber \\
&& \times \langle -\nu_a -\nu_b; -\bm{k}_{\textrm{i}}|\overline{\bm{M}}
  |-\nu_c  -\nu_d;-\bm{k}_{\textrm{f}}\rangle,
\label{eqA1}
\end{eqnarray}
where $\overline{\bm{M}}$ is the T-matrix for the inverse reaction.  
%According to the prescription , 
One can transform this relation to that for the amplitude $M_{\kappa}^{(K)}(s_{\textrm{i}} s_{\textrm{f}};\bm{k}_{\textrm{i}}  \bm{k}_{\textrm{f}})$ in Eq.\ (\ref{eq:Melemnts}) as
\begin{eqnarray}
&&M_{\kappa}^{(K)}(s_{\textrm{i}}s_{\textrm{f}};\bm{k}_{\textrm{i}}  \bm{k}_{\textrm{f}})
\nonumber \\
&=& (-)^{s_{\textrm{i}}+s_{\textrm{f}}-K} \overline{M}_{\kappa}^{(K)}(s_{\textrm{f}}s_{\textrm{i}};-\bm{k}_{\textrm{f}}, -\bm{k}_{\textrm{i}}).
\label{eqA2}
\end{eqnarray} 

We will transform the amplitude in the right-hand side of the above equation so that the direction of the momentum of the incident particle and that of the outgoing one are respectively same as those in the left-hand side amplitude.  
Since $\overline{\bm{M}}$ is the same as $\bm{M}$ for the elastic scattering, the transformation is described by the use of the rotation matrix $D$ \cite{Pr65} as
\begin{eqnarray}
&&\overline{M}_{\kappa}^{(K)}(s_{\textrm{f}} s_{\textrm{i}};-\bm{k}_{\textrm{f}}, -\bm{k}_{\textrm{i}})
\nonumber\\
 &=&\sum_{\kappa^{\prime}} D_{\kappa\kappa^{\prime}} (\pi-\theta, 0,\pi)
  M_{\kappa^{\prime}}^{(K)}(s_{\textrm{f}}s_{\textrm{i}};\bm{k}_{\textrm{i}}  \bm{k}_{\textrm{f}}).
\label{eqA3}
\end{eqnarray}
This leads to for $K=1$
\begin{equation}
M_1^{(1)}(01;\bm{k}_{\textrm{i}}  \bm{k}_{\textrm{f}})=-M_1^{(1)}(10;\bm{k}_{\textrm{i}}  \bm{k}_{\textrm{f}})
\label{eqA4}
\end{equation} 
and for $K=2$
\begin{eqnarray}
-\sqrt{\frac32} \frac{M_0^{(2)}(11;\bm{k}_{\textrm{i}}  \bm{k}_{\textrm{f}})}
  {\cos^2\theta-\sin^2\theta}&=&\frac{M_1^{(2)}(11;\bm{k}_{\textrm{i}}  \bm{k}_{\textrm{f}})}
  {\cos\theta\sin\theta}
\nonumber \\
&=&M_2^{(2)}(11;\bm{k}_{\textrm{i}}  \bm{k}_{\textrm{f}}),
\label{eqA5}
\end{eqnarray}
which provides
\begin{eqnarray}
&&\frac12 \left( \sqrt{\frac32}M_0^{(2)}(11;\bm{k}_{\textrm{i}}  \bm{k}_{\textrm{f}})
 - M_2^{(2)}(11;\bm{k}_{\textrm{i}}  \bm{k}_{\textrm{f}}) \right)
\nonumber\\
  &=& -\cot \theta M_1^{(2)}(11;\bm{k}_{\textrm{i}}  \bm{k}_{\textrm{f}}).
\label{eqA6}
\end{eqnarray}

Eqs.\ (\ref{eqA4}) and (\ref{eqA6}) are rewritten as
\begin{equation}
S_1=-S_2
\label{eqA7}
\end{equation}
and
\begin{equation}
\frac12 \left( \sqrt{\frac32}T_1-T_3 \right) = -\cot \theta T_2.
\label{A8}
\end{equation}

%%%%%%%%%%%%%%%%%%%%%%%%%%%%%%%%%%%%%%%%%%%%%%%%%%%%%%%%%%%%%%
\section{Depolarizations, polarization transfers, and spin correlations in general scattering between spin-1/2 particles}
\label{Sec:AppC}

In this appendix, the depolarizations $D_i^j(a)$ defined in Eq.\ (\ref{eq:Depol}), the polarization transfer coefficients $K_i^j(a \to d)$ in Eq.\ (\ref{eq:SpinTrns}), and the spin correlation coefficients $C_{ij}$ in Eq.\ (\ref{eq:SpinCorr}) are described in terms of the amplitudes in Eqs.\ (\ref{eq:Uj}), (\ref{eq:Sj}), and (\ref{eq:Tj}) for the case of general scattering between two spin-1/2 particles. 

The depolarizations of $a$ are described as
\begin{widetext}
\begin{eqnarray}
D_x^x(a) &=& \frac4{N_R} \textrm{Re} \left\{ 
 \frac1{2\sqrt3} \left(U_0+\frac1{\sqrt3}U_1\right)^{*} U_1 
 +\frac12\left(U_0-\frac1{\sqrt3}U_1\right)^{*} 
  \left(\frac1{\sqrt6}T_1-T_3 \right) \right.
\nonumber\\
& & \left. +\frac1{\sqrt2}\left(S_1-S_2\right)^{*}S_3 
  +\frac1{\sqrt2}\left(S_1+S_2\right)^{*}T_2 
  -\frac1{\sqrt6}T_1^{*}\left(\frac1{\sqrt6}T_1+T_3\right) \right\},
\label{eq:Dxx}
\end{eqnarray}
\begin{eqnarray}
D_y^y(a) &=& \frac4{N_R}\textrm{Re} \left\{ 
  \frac1{2\sqrt3} \left( U_0+\frac1{\sqrt3}U_1 \right)^{*} U_1 
  +\frac12 \left( U_0-\frac1{\sqrt3}U_1 \right)^{*} 
  \left( \frac1{\sqrt6}T_1+T_3 \right)  \right.
\nonumber\\
& & \left. -S_1^{*}S_2 +\frac12 |S_3|^2 
  -\frac1{\sqrt6}T_1^{*} \left( \frac1{\sqrt6}T_1-T_3 \right) 
   -\frac12|T_2|^2\right\},
\end{eqnarray} 
\begin{eqnarray}
D_z^z(a) &=& \frac4{N_R} \textrm{Re} \left\{ 
  \frac1{2\sqrt3} \left( U_0+\frac1{\sqrt3}U_1 \right)^{*} U_1 
  -\frac1{\sqrt6} \left( U_0-\frac1{\sqrt3}U_1 \right)^{*} T_1 \right.
\nonumber\\
& &\left. +\frac1{\sqrt2} \left( S_1-S_2 \right)^{*} S_3 
   -\frac1{\sqrt2} \left( S_1+S_2 \right)^{*} T_2 
   +\frac12 \left( \frac16 |T_1|^2-|T_3|^2 \right) \right\},
\end{eqnarray}
\begin{eqnarray}
D_x^z(a) &=& \frac4{N_R} \textrm{Re}\left\{ 
   \frac12 \left( U_0 + \frac1{\sqrt3} U_1 \right)^{*} S_3 
 - \frac{\sqrt2}{2\sqrt3} U_1^{*} \left( S_1 - S_2  \right)
 + \frac12 \left( U_0 - \frac1{\sqrt3} U_1 \right)^{*} T_2
 \right.
\nonumber\\
& & 
 - \frac1{\sqrt2}S_1^{*} \left( \frac1{\sqrt6}T_1 - T_3 \right)
 - \frac1{\sqrt3} S_2^{*} T_1
 + \frac12 S_3^{*} \left( \frac1{\sqrt6}T_1 + T_3 \right)
\nonumber\\
& & \left. -\frac12 T_2^{*}\left(\frac1{\sqrt6}T_1+T_3\right) \right\},
\end{eqnarray}
\begin{eqnarray}
D_z^x(a) &=& \frac4{N_R} \textrm{Re}\left\{ 
 - \frac12 \left( U_0 + \frac1{\sqrt3}U_1\right)^{*} S_3 
  +\frac{\sqrt2}{2\sqrt3} U_1^{*} \left( S_1- S_2 \right) 
  + \frac12 \left( U_0 - \frac1{\sqrt3} U_1 \right)^{*} T_2 \right.
\nonumber\\
& & 
 - \frac1{\sqrt3} S_1^{*} T_1
 - \frac1{\sqrt2}S_2^{*} \left( \frac1{\sqrt6}T_1 - T_3 \right)
 - \frac12 S_3^{*} \left( \frac1{\sqrt6}T_1 + T_3 \right)
\nonumber\\
& & \left.  -\frac12 T_2^{*}\left(\frac1{\sqrt6}T_1+T_3\right) \right\}.
\end{eqnarray}

The polarization transfer coefficients from $a$ to $d$ are described as
\begin{eqnarray}
K_x^x(a \to d) &=& \frac4{N_R} \textrm{Re}\left\{ 
 - \frac1{2\sqrt3} \left( U_0-\frac1{\sqrt3}U_1 \right)^{*} U_1 
 - \frac12 \left( U_0+\frac1{\sqrt3}U_1 \right)^{*} 
   \left( \frac1{\sqrt6}T_1-T_3 \right)  \right.
\nonumber\\
& & \left. 
  +\frac1{\sqrt2} \left( S_1 + S_2 \right)^{*} S_3 
  +\frac1{\sqrt2} \left( S_1 - S_2 \right)^{*} T_2 
  -\frac1{\sqrt6}T_1^{*} \left( \frac1{\sqrt6}T_1+T_3 \right) \right\},
\end{eqnarray}
\begin{eqnarray}
K_y^y(a \to d) &=& \frac4{N_R} \textrm{Re}\left\{ 
  -\frac1{2\sqrt3} \left( U_0-\frac1{\sqrt3}U_1 \right)^{*} U_1 
  -\frac12 \left( U_0+\frac1{\sqrt3}U_1 \right)^{*} 
      \left( \frac1{\sqrt6}T_1+T_3 \right)  \right.
\nonumber\\
& & \left. 
  + S_1^{*}S_2 + \frac12 |S_3|^2
  - \frac1{\sqrt6}T_1^{*} \left( \frac1{\sqrt6}T_1-T_3 \right) 
  - \frac12 |T_2|^2\right\},
\end{eqnarray}
\begin{eqnarray}
K_z^z(a \to d) &=& \frac4{N_R} \textrm{Re}\left\{ 
  - \frac1{2\sqrt3} \left( U_0-\frac1{\sqrt3}U_1 \right)^{*} U_1 
   +\frac1{\sqrt6} \left( U_0+\frac1{\sqrt3}U_1 \right)^{*} T_1 \right.
\nonumber\\
& & \left. 
  + \frac1{\sqrt2} \left( S_1 + S_2 \right)^{*} S_3 
  - \frac1{\sqrt2} \left( S_1 - S_2 \right)^{*} T_2 
  + \frac12 \left( \frac16|T_1|^2-|T_3|^2 \right) \right\},
\end{eqnarray}
\begin{eqnarray}
K_x^z(a \to d) &=& \frac4{N_R} \textrm{Re}\left\{
  -\frac12\left( U_0-\frac1{\sqrt3}U_1 \right)^{*} S_3 
  -\frac1{\sqrt6}U_1^{*} \left( S_1+ S_2 \right) 
  -\frac12\left( U_0 + \frac1{\sqrt3}U_1 \right)^{*} T_2 \right.
\nonumber\\
& &
  - \frac1{\sqrt2} S_1^{*} \left( \frac1{\sqrt6}T_1 - T_3 \right)
  + \frac1{\sqrt3} S_2^{*} T_1
  + \frac12 S_3^{*} \left( \frac1{\sqrt6}T_1 + T_3 \right)
\nonumber\\
& &\left. -\frac12 T_2^{*}\left(\frac1{\sqrt6}T_1+T_3\right) \right\},
\end{eqnarray}
\begin{eqnarray}
K_z^x(a \to d) &=& \frac4{N_R} \textrm{Re}\left\{ 
 \frac12 \left(U_0 - \frac1{\sqrt3}U_1 \right)^{*} S_3 
 +\frac1{\sqrt6}U_1^{*} \left( S_1 + S_2 \right) 
 -\frac12\left( U_0 + \frac1{\sqrt3}U_1 \right)^{*} T_2
 \right.
\nonumber\\
& & 
  - \frac1{\sqrt3} S_1^{*}T_1
  +\frac1{\sqrt2}S_2^{*} \left( \frac1{\sqrt6}T_1 - T_3 \right)
  -\frac12 S_3^{*} \left( \frac1{\sqrt6}T_1 + T_3 \right) 
\nonumber\\
& & \left. 
 -\frac12 T_2^{*}\left(\frac1{\sqrt6}T_1+T_3\right) \right\}.
\end{eqnarray}

The spin correlation coefficients are described as
\begin{eqnarray}
C_{xx} &=& \frac1{N_R} \textrm{Re}\left\{ 
 -|U_0|^2+\frac13|U_1|^2 
 -\frac4{\sqrt3}U_1^{*} \left( \frac1{\sqrt6}T_1-T_3 \right)  \right.
\nonumber\\
& & \left. +2 \left( -|S_1|^2+|S_2|^2 \right)  -4S_3^{*}T_2 +\frac4{\sqrt6}T_1^{*} \left( \frac1{\sqrt6}T_1+T_3 \right) \right\},
\end{eqnarray}
\begin{eqnarray}
C_{yy} &=& \frac1{N_R} \textrm{Re}\left\{ 
 -|U_0|^2+\frac13|U_1|^2 
 -\frac4{\sqrt3}U_1^{*} \left( \frac1{\sqrt6}T_1+T_3 \right)  \right.
\nonumber\\
& & \left. -2 \left( |S_1|^2+|S_2|^2-|S_3|^2 \right)  +\frac4{\sqrt6}T_1^{*} \left( \frac1{\sqrt6}T_1-T_3 \right) +2|T_2|^2\right\},
\end{eqnarray}
\begin{eqnarray}
C_{zz} &=& \frac1{N_R} \textrm{Re}\left\{ 
  -|U_0|^2+\frac13|U_1|^2 
  +\frac{4\sqrt2}3U_1^{*}T_1 \right.
\nonumber\\
& & \left. +2 \left( -|S_1|^2+|S_2|^2 \right)  +4S_3^{*}T_2 -\frac13|T_1|^2+2|T_3|^2\right\},
\end{eqnarray}
\begin{eqnarray}
C_{xz} &=& \frac4{N_R} \textrm{Re}\left\{ 
 -\frac1{\sqrt2}U_0^{*}S_2 
  -\frac1{\sqrt3}U_1^{*} \left( \frac1{\sqrt2}S_1 +T_2 \right)  \right.
\nonumber\\
& & 
 - \frac1{\sqrt2} S_1^{*} \left( \frac1{\sqrt6} T_1 + T_3 \right)
 + \frac12 S_3^{*} \left( \frac3{\sqrt6}T_1 - T_3 \right)
\nonumber\\
& & \left. 
 + \frac12 T_2^{*}\left(\frac1{\sqrt6}T_1+T_3\right) \right\}.
\end{eqnarray}
\begin{eqnarray}
C_{zx} &=& \frac4{N_R} \textrm{Re}\left\{ 
 \frac1{\sqrt2}U_0^{*}S_2 
 +\frac1{\sqrt3}U_1^{*} \left( \frac1{\sqrt2}S_1-T_2 \right)  \right.
\nonumber\\
& & 
 + \frac1{\sqrt2} S_1^{*} \left( \frac1{\sqrt6} T_1 + T_3 \right)
 + \frac12 S_3^{*} \left( \frac3{\sqrt6}T_1 - T_3 \right)
\nonumber\\
& & \left. 
  +\frac12 T_2^{*}\left(\frac1{\sqrt6}T_1+T_3\right) \right\}.
\end{eqnarray}
For the combinations of $i$ and $j$, $(ij)=(xy), (yx), (yz), (zy)$, the quantities $D_i^j$, $K_i^j$, and $C_{ij}$ automatically vanish.
\end{widetext}

%%%%%%%%%%%%%%%%%%%%%%%%%%%%%%%%%%%%%%%

\end{document}